\documentclass{sigchi-ext}
\usepackage[T1]{fontenc}
\usepackage{textcomp}
\usepackage[scaled=.92]{helvet} 
\usepackage{graphicx} 
\usepackage{balance}  
\usepackage{booktabs} 
\usepackage{ccicons}  
\usepackage{ragged2e} 

\def\plaintitle{SIGCHI Extended Abstracts Sample File: Note Initial
  Caps} 
\def\emptyauthor{}
\def\plainkeywords{Authors' choice; of terms; separated; by
  semicolons; include commas, within terms only; required.}

\title{``Please enter your PIN'' - On the Risk of Bypass Attacks on Biometric Authentication on Mobile Devices}

\numberofauthors{3}

\author{%
  \alignauthor{%
    \textbf{Christian Tiefenau}\\
    \affaddr{University of Bonn} \\
    \affaddr{Bonn, Germany} \\
    \email{tiefenau@cs.uni-bonn.de} 
  }
  \alignauthor{%
    \textbf{Maximilian H{\"a}ring}\\
    \affaddr{Fraunhofer FKIE}\\
    \affaddr{Bonn, Germany}\\
    \email{haering@cs.uni-bonn.de} 
  } \vfil 
  \alignauthor{%
    \textbf{Mohamed Khamis}\\
    \affaddr{University of Glasgow}\\
    \affaddr{Glasgow, UK}\\
    \email{mohamed.khamis@\\glasgow.ac.uk} 
  } 
  \alignauthor{%
    \textbf{Emanuel von Zezschwitz}\\
    \affaddr{University of Bonn, Fraunhofer FKIE}\\
    \affaddr{Bonn, Germany}\\
    \email{zezschwitz@cs.uni-bonn.de} 
  }
}

\definecolor{linkColor}{RGB}{6,125,233}
\hypersetup{%
  pdftitle={\plaintitle},
  pdfauthor={\emptyauthor},
  pdfkeywords={\plainkeywords},
  bookmarksnumbered,
  pdfstartview={FitH},
  colorlinks,
  citecolor=black,
  filecolor=black,
  linkcolor=black,
  urlcolor=linkColor,
  breaklinks=true,
}

\begin{document}
\begin{marginfigure}[8pc]
  \begin{minipage}{\marginparwidth}
    \centering
    \includegraphics[width=0.9\marginparwidth]{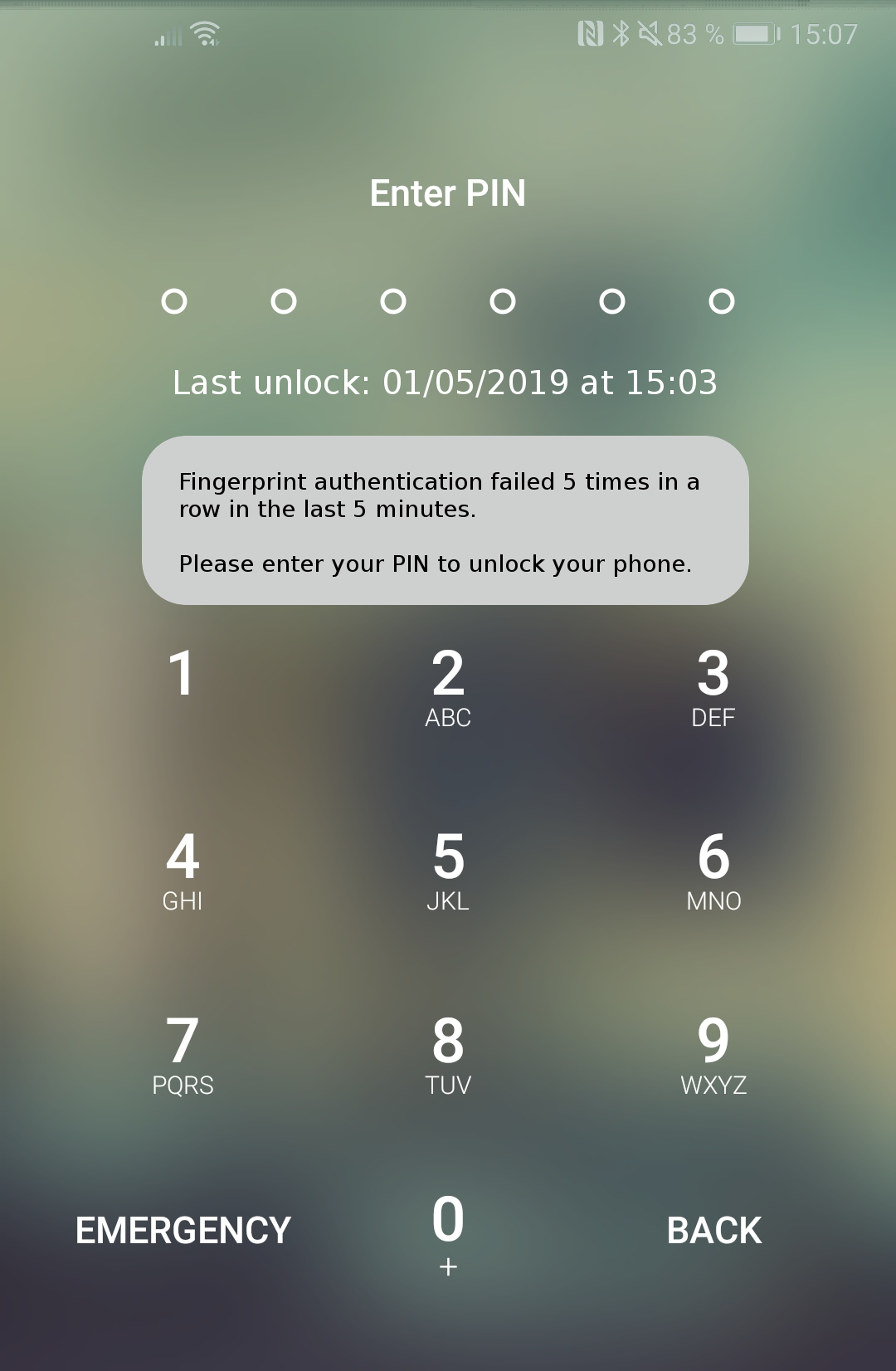}
    \caption{Mockup of an Android lock screen using a reactive security message that reports the last unlock time. In addition, the reason why the user has to enter the PIN instead of using the fingerprint is displayed.}
    \label{fig:android_mockup_full}
  \end{minipage}
\end{marginfigure}
\copyrightinfo{Copyright is held by the author/owner. Permission to make digital or hard copies of all or part of this work for personal or classroom use is granted without fee. Poster presented at the 15th Symposium on Usable Privacy and Security (SOUPS 2019).}

\maketitle

\RaggedRight{} 

\begin{abstract} 
Nowadays, most mobile devices support biometric authentication schemes like fingerprint or face unlock. However, these probabilistic mechanisms can only be activated in combination with a second alternative factor, usually knowledge-based authentication. In this paper, we show that this aspect can be exploited in a bypass attack. In this bypass attack, the attacker forces the user to ``bypass'' the biometric authentication by, for example, resetting the phone. 
This forces the user to enter an easy-to-observe passcode instead. We present the threat model and provide preliminary results of an online survey. Based on our results, we discuss potential countermeasures. We conclude that better feedback design and security-optimized fallback mechanisms can help further improve the overall security of mobile unlock mechanisms while preserving usability. 
\end{abstract}

\section{Introduction and Background}
Mobile devices store a lot of potentially sensitive information, e.g., in the form of e-mails, pictures, or contact data. A lock screen, which requires user authentication, is often the last barrier of protection from unauthorized access. 
While knowledge-based approaches (e.g., PIN) have been the quasi-standard for many years, biometric approaches (e.g., fingerprint) are increasingly employed as a more usable alternative. This can be explained by recent research which shows that users unlock their mobile devices several times an hour \cite{harbach2014sa}, which underlines the importance of usability in terms of speed and error rate.
Indeed, in comparison to their to knowledge-based counterparts, biometric authentication schemes are faster and easier to use \cite{Bhagavatula2015}. 
Biometric authentication is also superior in terms of security as it resists many attacks that knowledge-based schemes are vulnerable to, such as observation  \cite{vonZezschwitz:2015:EDB:2702123.2702202}, smudge \cite{Aviv:2010:SAS:1925004.1925009} and thermal attacks \cite{Abdelrahman:2017:SCU:3025453.3025461}. 

\begin{marginfigure}
    \includegraphics[width=0.9\marginparwidth]{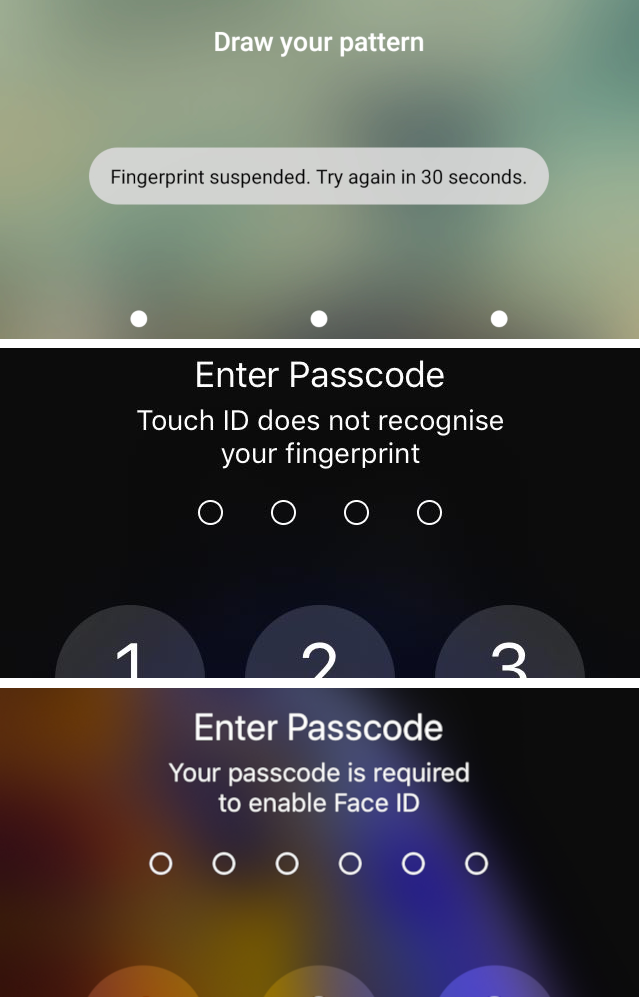}
    \caption{Notification of Android 8 and iOS 12.1.2 on an iPhone 5s and X if the fingerprint was not correctly identified 5 times in a row or the device was unable to identify the users' face. We assume that a more detailed message (e.g., ``fingerprint authentication failed 5 times in a row in the last 5 minutes'') could improve both the usability of the system and the security consciousness of the user.}
    \label{fig:lockscreen_warning}
\end{marginfigure}

Nevertheless, biometric authentication does not come without limitations. Previous work showed that biometric authentication can be vulnerable to sophisticated camera-based smudge attacks, and sometimes suffers from recognition errors when the user's features are distorted or occluded (e.g., using TouchID with dirty or wet fingers, or using FaceID when only part of the face is visible). 

We identify an additional vulnerability in biometric authentication that has not been investigated in prior research. 
Namely, because of their probabilistic nature, mobile biometric authentication always requires a second deterministic mode, which in turn reduces security. 
For example, Android and iOS require users to set a knowledge-based password (e.g., PIN or Lock Pattern) whenever they choose biometric authentication as the main unlocking mechanism\footnote{iOS: \url{https://www.apple.com/business/site/docs/iOS_Security_Guide.pdf}\\Android:  \url{https://source.android.com/security/encryption/full-disk.html}}. 
We argue that this is a security issue, as lock screens that support multiple authentication modes can only be as secure as the weakest mechanism provided.

In this paper, we show that the current lock screen implementations are still prone to observation attacks, as users can easily be tricked into using the knowledge-based factor. 
For example, an attacker could repeatedly push their finger against a user's fingerprint sensor to force the system to require a knowledge-based password, which in turn can be easily uncovered via observation, thermal or smudge attacks.  
Therefore, we argue that there is little need for educated attacks on biometrics, as long as the provided alternative mode (e.g., PIN) is still vulnerable. 

We discuss our threat model and present several attack vectors which exploit the human factor in combination with flawed interface design (i.e., the lack of feedback). We demonstrate that current mobile phones can be quickly manipulated in a way that \textit{bypasses} biometric authentication and thereby forcing users to enter their passcode. The results of an online survey (N=200) indicate, that most users do not question switches from biometric to knowledge-based authentication, which suggests that users are vulnerable to said threat. Based on our results, we discuss potential countermeasures and illustrate directions for future work. We propose small changes to the user interface (i.e., provide feedback on mode switches) which could prevent the \textit{bypass attack}. Furthermore, we highlight the need to revisit previously discussed security-optimized (e.g., observation-resistant) authentication mechanisms in the light of the new context (i.e., fallback authentication).

\section{Threat Model}
In our threat model, 
we assume that the attacker is in the vicinity of the victim and has short-term access to the victim's mobile device. The victim's device is configured with a biometric lock screen such as fingerprint, as a fallback mechanism, the victim has configured PIN. 

The attacker does not know the victim's PIN. Once the victim's device is unattended, the attacker prepares the device in a way that a PIN-based authentication will be required for the next unlock. This can be done by triggering 5 unsuccessful biometric authentication attempts\footnote{\url{https://support.apple.com/en-us/HT204587}}. 
As soon as the victim tries to unlock their phone, the system asks for the PIN. 
Without questioning the reason, the victim provides their PIN-code. 
The attacker then observes the entered PIN, gets hold of the device (e.g., when the user leaves it unattended), and then unlocks the device after switching to the fallback screen. 
In this scenario, there is no need to break the biometric factor. 
Since the device does not provide any information about switches between factors, the victim remains unaware that someone tampered with their phone. 

\section{Practical Execution of the Bypass Attack}

In contrast to previous elaborated attacks on biometric sensors, we investigated 
non-detectable low-cost approaches which force the victim to bypass the biometric unlock and enter the fallback authentication code. Based on the review of the currently deployed hardware and current lock screen implementations, we outline software-based and hardware-based attacks. 

\begin{marginfigure}[-32pc]
    \includegraphics[width=\marginparwidth]{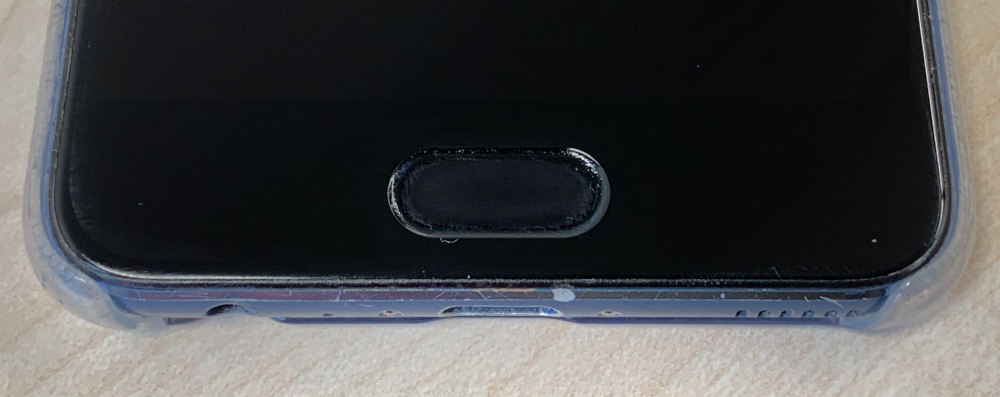}
    \caption{Home button covered with lip balm to prevent the fingerprint sensor from recognizing the users' fingerprint.}
    \label{fig:fingerprint_sensor}
\end{marginfigure}

{\tiny
\begin{margintable} [-12pc]
    \begin{tabular}{l | c c | c c}
      & \multicolumn{2}{c|}{\small \textbf{iPhone}} & \multicolumn{2}{c}{\small \textbf{Android}}\\
      & prim. & sec. & prim. & sec. \\
      \toprule
      PIN & 27 & 51 & 30 & 56 \\
      Pattern & 19 & 31 & 0 & 7 \\
      Password & 6 & 32 & 8 & 22 \\
      Fingerprint & 42 & 21 & 55 & 23\\
      Face unlock & 2 & 5 & 6 & 9\\
      Other & 4 & 4 & 1 & 2\\
      \bottomrule
      Total & 100 & 144 & 100 & 119 \\
    \end{tabular}
    \caption{The distribution of both device-groups divided by the primary and secondary method as reported by the participants.}
    \label{tab:distribution_pim_sec_method}
\end{margintable}
}

\begin{marginfigure}
    \includegraphics[width=\marginparwidth]{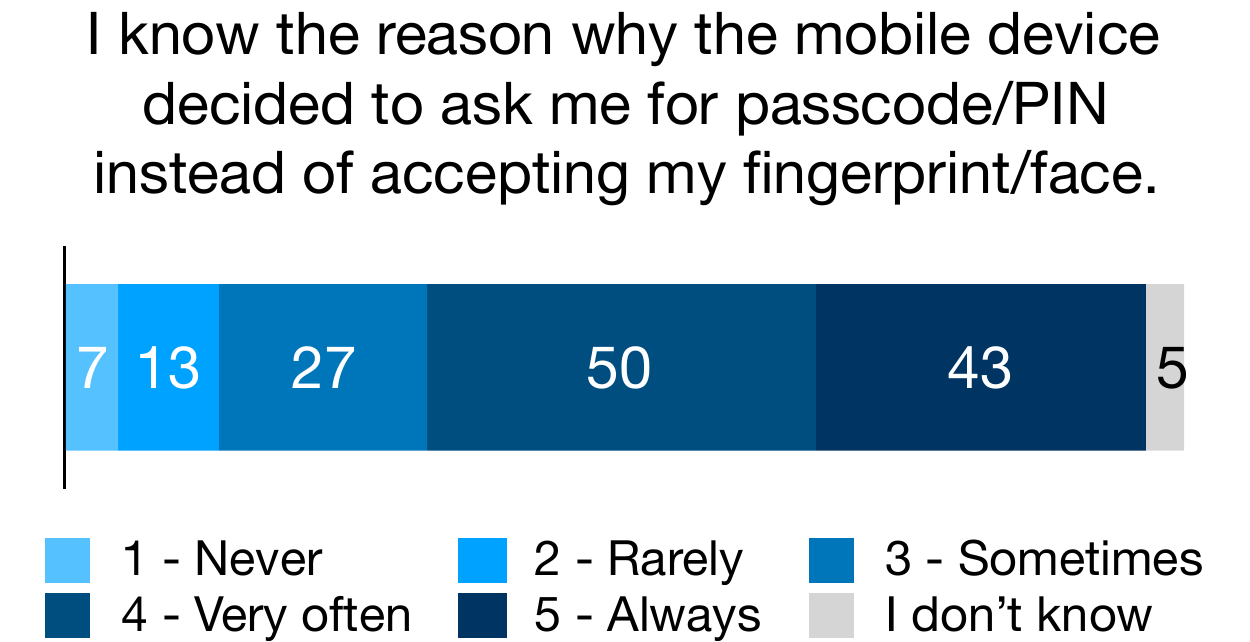}
\end{marginfigure}
\begin{marginfigure}
    \includegraphics[width=\marginparwidth]{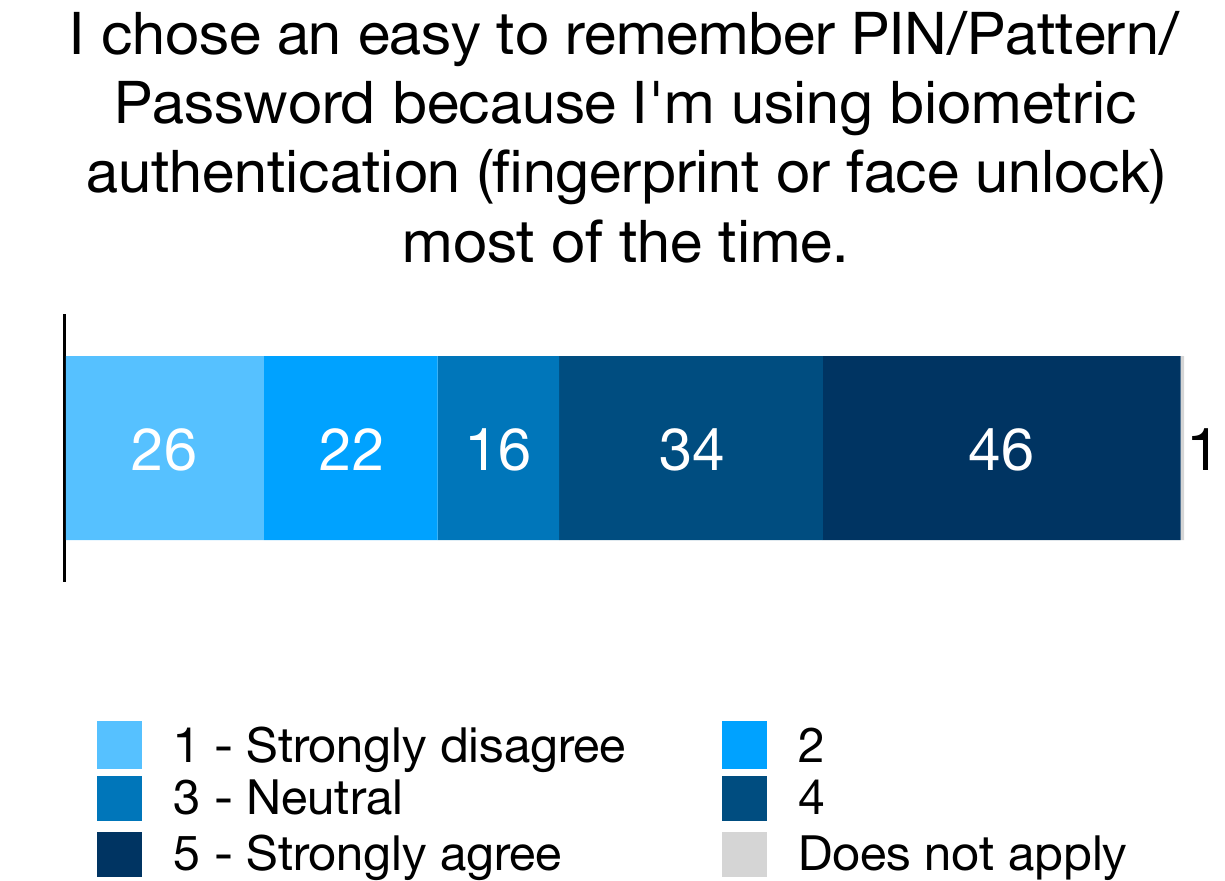}
    \caption{Likert scale answers of the survey.}
    \label{fig:likertscale}
\end{marginfigure}

\subsection{Software-based Attacks}
Software-based attacks exploit the implementation of current lock screens. For example, current iPhones require the passcode 1) after reboot, and 2) after five failed biometric attempts. This aspect can be exploited by an attacker to actively require users to enter their passcode; the attacker can switch off / restart the device or perform five failed attempts using their own finger or face (depending on the used mechanism). As illustrated in Figure \ref{fig:lockscreen_warning}, current mobile devices provide only limited feedback on mode switches. Therefore, users are likely to remain unaware of the attack.

\subsection{Hardware-based Attacks}
Attackers can actively manipulate the biometric sensor. We performed a number of experiments with the iPhone 5s and explored different ways. 
We found that applying a) greasy lip balm or b) some drops of water prevents the sensors from working. While this method may be more eye-catching, it is an easy and direct way to force users to enter their passcode.

\section{User Behavior and Risk Perception}
To gather preliminary results on the risk of the presented threat model, we used Amazons MTurk to survey 200 US-citizens (100 Android, 100 iPhone) about their unlock mechanisms. The survey took about 5 minutes and was compensated with 0.75\$. We were interested in the participants' understanding of mode switches. This means, if they understand why their device is requiring a passcode. In addition, we asked how often they leave their smartphone unattended and if the use of biometrics affects the choice of passcodes.

Table \ref{tab:distribution_pim_sec_method} provides self-reported data on primary and secondary authentication. In the survey, we asked the participants which methods they ``usually used to unlock'' (primary) and ``all other activated unlock mechanisms'' they have (secondary). 145 participants (61 iPhone, 84 Android) reported to use fingerprint or face unlock as \emph{primary} or \emph{secondary} method. Please note that in this particular case ``primary'' and ``secondary'' was related to the participant's perceived usage of the respective method. 

We used a 5-point scale (\emph{1 - Never} to \emph{5 - Always}) to ask biometric users ``how often they know the reason why their mobile device asked them for their PIN or passcode instead of accepting their biometric login''. 97 of 145 (~67\%), as seen in Figure~\ref{fig:likertscale} reported, that they do not always understand the reasons of mode switches, seven iPhone users acknowledged that they can never figure out the reason. This result indicates that users are likely to remain unaware of malicious mode switches. In addition, we asked to what extent the participants agree to the statement \emph{``I chose an easy to remember PIN/Pattern/Password because I'm using biometric authentication (fingerprint or face unlock) most of the time.''}. 
As seen in Figure \ref{fig:likertscale}, we found a trend indicating that users select weaker passwords when they use biometric authentication. 80 of 145 (~55\%) participants agreed that they deliberately chose an easy password. We asked how often the participants leave their phone unattended when e.g., going to the rest rooms. 135 of 200 (67.5\%) reported that they occasionally or often do this, which makes short-term access for the attacker possible. 

\section{Potential Countermeasures}
Based on our preliminary results, we propose to investigate the effectiveness of the following countermeasures. 

\begin{marginfigure}
\vspace{-20mm}
\begin{flushleft}
{\tiny
    \texttt{@@@@@@@@@@@@@@@@@@@@@@@@@@@@@@@@@@@@@@ \\
WARNING:POSSIBLE DNS SPOOFING DETECTED \\
@@@@@@@@@@@@@@@@@@@@@@@@@@@@@@@@@@@@@@ \\
The ECDSA host key for server has  \\
changed, and the key for the  \\
corresponding IP address 192.168.0.1  \\
is unchanged. This could either mean  \\
that DNS SPOOFING is happening or the \\
IP address for the host and its host \\
key have changed at the same time.\\
}}
\end{flushleft}
 \caption{Warning message of SSH when the IP-address of a known server has changed. Similarly, we recommend showing users a warning message when biometric authentication fails to improve their security consciousness. }
    \label{fig:sshsample}
\end{marginfigure}

\subsection{Improved Interface Design}
Most users do not question the fact that a passcode is required. Our preliminary results indicate that the system state of current lock screens is not always easy to assess. We argue that future work needs to investigate if the unlock screen can be improved in a way that better communicates the current system state. Following the usability heuristics by Jakob Nielsen \cite{nielsen199510}, we recommend to provide more information about mode switches between primary and secondary authentication. Figure \ref{fig:android_mockup_full} shows an example of how such feedback could look like. We would recommend to inform the user about the last unlock event, maybe only when it seems necessary, e.g., determined by uncharacteristic behavior. We assume that this information could help the user to discover ongoing security breaches (cf., threat model). Similar reactive security measures are already in use in OpenSSH (see Figure \ref{fig:sshsample}) or in online banking platforms.

\subsection{Improved Knowledge-based Authentication}
Our preliminary findings indicate that biometric authentication schemes are often used as the primary authentication method. As a consequence, the knowledge-based approach is used less frequently. We argue that this aspect changes the requirements for feasible knowledge-based authentication mechanisms. We assume that the new context of use renders efficiency (i.e., input time) less important and makes memorability and effectiveness more important. Previous research has proposed a plethora of knowledge-based authentication mechanisms which were optimized to resist different types of side-channel attacks like observation attacks \cite{Eiband:2017:USS:3025453.3025636}, smudge attacks \cite{Aviv:2010:SAS:1925004.1925009} or thermal attacks \cite{Abdelrahman:2017:SCU:3025453.3025461}. While these mechanisms (e.g., \cite{Bianchi2012409, gugenheimer2015colorsnakes}) were shown to be more secure, they were often too slow to be used as a primary (i.e., most used) mobile authentication method. We recommend to review previous work in the light of the new use case (i.e., as a fallback authentication). We assume that using such security-optimized authentication schemes becomes feasible in the new context and would improve the overall security of the device.

\section{Conclusion and Future Work}
In this paper, we discussed a new threat model, a bypass attack. By performing a bypass attack, the attacker is forcing the user to ``bypass'' biometric authentication and to use an easy-to-observe knowledge-based mode instead. Based on the preliminary results of an online survey, we discussed potential countermeasures which need to be investigated in future work. We argued that the overall security of the lock screens could be improved by 1) providing more feedback to the user and by 2) deploying security-optimized (e.g., observation-resistant) knowledge-based mechanisms. 

\balance{} 

\bibliographystyle{SIGCHI-Reference-Format}
\bibliography{sample}

\end{document}